\documentclass[conference]{IEEEtran}

\usepackage{xcolor}
\usepackage{gensymb}
\usepackage{subcaption}
\usepackage{graphicx} 
\usepackage{amsmath}
\usepackage{amssymb}
\usepackage{tabularx,booktabs}
\usepackage{diagbox}
\usepackage[numbers]{natbib}
\usepackage[ruled, vlined]{algorithm2e}
\usepackage[pagebackref=true,breaklinks=true,colorlinks,bookmarks=false]{hyperref}

\newcolumntype{C}{>{\centering\arraybackslash}X}
\newcommand*{\rom}[1]{\expandafter\@slowromancap\romannumeral #1@}

\begin{document}
\title{Dynamic Adversarial Patch for Evading Object Detection Models}

\author{
        \textbf{\large Shahar Hoory, Tzvika Shapira, Asaf Shabtai, Yuval Elovici}
    
        \\Department of Software and Information Systems Engineering,
        \\Ben-Gurion University of the Negev, Be’er Sheva, Israel
        \\\{hoorysha,avrahamt\}@post.bgu.ac.il, \{shabtaia,elovici\}@bgu.ac.il
        
}

\maketitle

\begin{abstract}
Neural networks (NNs) have been widely adopted for detecting objects in images and video streams.
Recent research shows that NN models used for classification are vulnerable to adversarial evasion attacks, in which minimal changes to the input can result in misclassification.
YOLO and Fast R-CNN are examples of object detectors that have been shown to be vulnerable to adversarial attacks.
Most of the existing real-world adversarial attacks against object detectors use an adversarial patch which is attached to the target object (e.g., a carefully crafted sticker placed on a stop sign).
This method may not be robust to changes in the camera's location relative to the target object; in addition, it may not work well when applied to nonplanar objects such as cars.
In this study, we present an innovative attack method against object detectors applied in a real-world setup that addresses some of the limitations of existing attacks - \textit{the Dynamic Adversarial Patch}.
Our method uses dynamic adversarial patches which are placed at multiple predetermined locations on a target object.
An adversarial learning algorithm is applied in order to generate the patches used. 
The dynamic attack is implemented by switching between optimized patches dynamically, according to the camera's position (i.e., the object detection system's position).
In order to demonstrate our attack in a real-world setup, we implemented the patches by attaching flat screens to the target object; the screens are used to present the patches and switch between them, depending on the current camera location.
Thus, the attack is dynamic and adjusts itself to the situation to achieve optimal results.
We evaluated our dynamic patch approach by attacking the YOLOv2 object detector with a car as the target object and succeeded in misleading it in up to 90\% of the video frames when filming the car from a wide viewing angle range.
We improved the attack by generating patches that consider the semantic distance between the target object and its classification.
We also examined the attack's transferability among different car models and were able to mislead the detector 71\% of the time.
\end{abstract}

\section{Introduction}
Deep neural networks (DNNs) are widely used in many domains.
One of them is computer vision, in which DNNs are employed for object detection
~\cite{redmon2016you,redmon2017yolo9000,redmon2018yolov3,bochkovskiy2020yolov4,girshick2014rich,girshick2015fast,ren2015faster,he2017mask}.
Object detection models assist in performing various tasks, such as autonomous driving~\cite{chen2016monocular, chen2017multi, wu2017squeezedet}, security video analytics~\cite{anjum2016video}, object tracking~\cite{ning2017spatially}, and more.

Recent studies have shown that DNNs are vulnerable to adversarial attacks~\cite{szegedy2013intriguing,carlini2017towards,goodfellow2014explaining,papernot2016limitations,moosavi2017universal}.
Adversarial attacks can be classified as either poisoning attacks or evasion attacks~\cite{chakraborty2018adversarial}.
In \textit{poisoning attacks}~\cite{biggio2012poisoning}, the attacker influences the training phase of the model by injecting malicious samples into the training set.
In \textit{evasion attacks}~\cite{biggio2013evasion}, the attacker tries to evade the system by crafting perturbed examples that lead to incorrect classification during the inference phase.

State-of-the-art object detection models are based on DNNs~\cite{redmon2016you,redmon2017yolo9000,redmon2018yolov3,bochkovskiy2020yolov4,girshick2014rich,girshick2015fast,ren2015faster,he2017mask}, and therefore they are also vulnerable to adversarial attacks.
Since these models are incorporated in many safety-critical systems, such as advanced driver-assistance systems (ADAS)~\cite{lillicrap2015continuous, zhou2018voxelnet, bojarski2016end}, a successful adversarial attack may result in severe consequences.

Initial adversarial attacks on object detection models were digital attacks, i.e., implemented by adding a small perturbation to an image file such that its classification by the target model would change~\cite{xie2017adversarial, eykholt2018robust}. 
Recently, physical end-to-end attacks have been demonstrated. 
This was achieved by modifying the scene, usually by using a patch which is placed on the object before an image is taken and processed by a DNN model~\cite{brown1712adversarial,eykholt2018physical,zhao2018seeing,thys2019fooling}; for example, adversarial stickers have been attached to stop signs to prevent their detection~\cite{eykholt2018physical,zhao2018seeing}.
Another example of a real-world adversarial attack is creating an adversarially perturbed stop sign that is misdetected by an object detector~\cite{chen2018shapeshifter}.

Recently, real-world adversarial attacks aimed at enabling a nonplanar object (i.e., 3D object), such as a person, to evade detection by an object detector were presented by \citet{thys2019fooling} and \citet{huang2020universal}; such attacks are much more difficult to perform than attacks on 2D objects. 
Despite the success of the abovementioned attacks, the studies had limitations.
\citet{thys2019fooling} did not evaluate the effect of the viewing angle on the attack's success, and the real-world experiments of \citet{huang2020universal} showed that as the absolute value of the angle between the object and the camera increases, the attack becomes less successful.

\citet{lu2017no} evaluated the robustness of state-of-the-art, real-world attacks (in the form of a printed \textit{fixed and static} patch) on object detection models.
Their experiments showed that the distance between the camera and the target object affects the attack success.
The authors also conducted a limited experiment showing the impact of different view angles on the attack success. 
Consequently, they raised the following important question: ``Can one construct examples that are adversarial for many or most viewing conditions?''

In this work, we discuss the limitations of attacks aimed at object detectors when the target objects are \textit{large} and \textit{nonplanar} and propose a novel attack method against object detectors in the real world: \textbf{the dynamic adversarial patch.} 
We specifically explore the effect of the filming angle on the attack success and present an attack method which addresses the challenge raised by~\citet{lu2017no}, by adapting the adversarial patches dynamically according to the current camera location, thus increasing the attack's robustness.
To demonstrate our attack method, we use a \textbf{car} as our target object and attach flat screens to it in order to add the dynamic patches to the scene. 
The adversarial patch presented on the screen is changed according to the location of the camera.
The goal of the dynamic adversarial patches is either to cause the object detector to fail to detect the car or to detect and misclassify a car, classifying it as any other object (i.e., an untargeted attack).
In addition, since for a large, nonplanner object, like a car, placing a patch in one location may not be sufficient (since at some filming positions and angles the patch may not be visible to the camera), we opted to apply our dynamic patches in multiple places on the object (using multiple screens).
We further improve our attack so that the target object (i.e., car) will be misclassified as a \textit{semantically-unrelated} object (as opposed to a semantically-related object, such as a bus or truck) by modifying the loss function of the attack according to the semantic relation between the objects.
Our method has two unique advantages that distinguish it from existing adversarial attacks: \textit{i)} it can interfere with the detection of large, nonplanar objects such as cars, and \textit{ii)} it is effective for a wide range of view angles. We consider two threat model scenarios in which our attack is relevant: when the target object is in motion and when it is stationary.

In order to evaluate our proposed method, we conducted a set of experiments that evaluate the effectiveness of the attack on the YOLOv2 object detector~\cite{redmon2017yolo9000}.
The results of our experiments show that the proposed dynamic adversarial patch method is capable of both preventing YOLOv2 from detecting a car and causing the object detector to misclassify it when filming from a wide range of angles, achieving a success rate of up to 90\% of video frames in which the model didn't detect car.
In addition, transferability experiments show that the attack is transferable among different car models with a success rate of up to 71\% 
Finally, we show that a patch aimed at causing the detector to classify the object as a semantically-unrelated class is feasible, obtaining a success rate of up to 72\% in a wide range of viewing angles.

We summarize the contributions of this paper as follows:
\begin{itemize}
 \item We present the challenges associated with attacking object detectors when targeting large, nonplanar objects.
 \item We propose a novel adversarial attack technique that uses multiple dynamic adversarial patches attached to the target object.
 The attack is implemented by switching between pregenerated patches which are optimized for the current camera location.
 \item We evaluate the method in a real-world setting using a wide range of filming angles when the target object is car.
 \item We demonstrate the transferability of the attack when the attack is crafted on one car model and applied to a different car model.
  \item We improve the attack by considering the semantic distance between the target object and its classification.
\end{itemize}

\section{\label{sec_relatedWork}Related Work}
\subsection{Adversarial Attacks}
\citet{szegedy2013intriguing} were the first to present adversarial attacks against neural networks.
They discovered that deep neural networks are vulnerable to imperceptible perturbations of the input that can result in changes to the output and thus cause misclassification.
The perturbed inputs are termed ``adversarial examples."
These attacks can be targeted or untargeted and are often described as an optimization problem: given an input \(x\), a model \(f\colon \mathbb{R}^m \rightarrow \{1,\dots,k\}\), an actual label y where \(y=f(x)\), and an attack target label \(l \neq y\), the goal is to find a perturbation \(r\) that minimizes \(\|r\|\) subject to \(f(x+r)=l\) for \textit{targeted} attacks, and subject to \(f(x+r) \neq y\) for \textit{untargeted} attacks.

Subsequent studies expanded upon this and proposed other properties of neural networks which explain adversarial examples. 
\citet{goodfellow2014explaining} argued that the linear nature of models is the primary reason for their vulnerability to adversarial attacks, and based on this claim, the authors suggested the fast gradient sign method (FGSM) attack in which the perturbation is computed as follows:
\[\eta = sign (\nabla_x J(\theta,x,y))\]
where \(\theta\) are the model's parameters, \(x\) is the input, \(y\) is the input's label, and \(J\) is the cost function used to train the model.

Since those foundational studies, many papers have addressed this area and presented diverse attack methods. 
\citet{moosavi2016deepfool} introduced DeepFool, an iterative untargeted adversarial attack, which is based on linearization of the model, as in~\cite{goodfellow2014explaining}.
The attack creates minimal perturbations that are sufficient for misleading the model.
Furthermore, they showed that adversarial fine-tuning (training a model with perturbed input as a part of the training set) significantly increases the model's robustness to adversarial attacks.

In 2017, \citet{carlini2017towards} presented powerful attacks that have been shown to be more effective than existing attacks.
Each attack is limited to a different distance metric: \(l_0\), \( l_2\),  and  \(l_\infty\).
The attacks obtain their input from the logits layer instead of the softmax layer and use a different loss function than that used by the model.
\citeauthor{carlini2017towards}'s attacks~\cite{carlini2017towards} were also successful against the distillation defense method proposed by \citet{papernot2016distillation}.

While most attacks are based on the ability to backpropagate the gradients from the network's output, \citet{papernot2016limitations} proposed a different attack: the Jacobian-based saliency map attack (JSMA), which is based on the forward derivatives of the model's learned function.
In the JSMA, a saliency map is created, which determines the features in the input space to be affected by the perturbation in order to decrease the network's accuracy in the most effective way.
This attack ensures that the perturbation focuses on those features and changes them until the adversarial example is misclassified.

\subsection{Adversarial Attacks in the Real World}
There are two common attacks in the area of adversarial attacks on visually-related models (such as facial recognition, image classification, and object detection): digital space attacks and real-world attacks.
In contrast to attacks in the digital space (i.e., directly changing the pixels of the input image), attacks in the real world require changing the actual object in the real world (e.g., by physically adding a perturbation to it) before a camera captures the scene.
Although the abovementioned attacks have achieved high success rates, they are not applicable to the real world, where there is noise from the environment.
For example, a perturbation added to an input image which is used to attack an image classifier may not be successful in the real world, because the camera taking the photos is accompanied by noise that affects the pixels, along with the lighting conditions and other factors, all of which must be taken into account when performing the attack.

\citet{kurakin2016adversarial} presented attacks based on FGSM that are applicable to the real world.
They showed that photos of printed adversarial images obtained from a cell phone camera can fool image classification models. However, they don't describe a process for generating the noise with respect to real world constraints.

\citet{athalye2017synthesizing} were the first to show the existence of real-world perturbations that are robust to real world constraints such as noise, different lighting conditions, and camera angles. 
They applied their attack to physical 3D objects printed using 3D printing and created the first adversarial object capable of fooling classifiers in the physical world. 
The authors also developed the expectation over transformation (EOT) algorithm, which produces perturbations that are adversarial over an entire distribution of transformations and hence can fool classifiers in real-world settings.
\citet{brown1712adversarial} also used the EOT method and developed
a printable robust adversarial patch which is added to a scene, photographed, and fed into a classifier which then misclassified the image.
The adversarial patch method has formed the basics to multiple  real world attacks (presented in subsection~\ref{subsec_realWorldAttacks}).
Our attack is also inspired by this method, however, we implement an adaptive patch (that changes according to the camera view point) and the patch is presented using flat screens instead of a printed sticker.

\citet{sharif2016accessorize} showed a potential usage of adversarial attacks in the real world in the form of adversarial accessories, presenting a pair of adversarial glasses that fool facial recognition systems in the real world.

\subsection{Object Detection}
In recent years, there has been great progress in the field of object detection. 
There are two main object detection approaches: region proposal-based models and one-stage models~\cite{liu2020deep}.
Region proposal models typically use a preprocessing stage to generate areas in the image called \textit{region proposals}.
Features are extracted from each region proposal, and a classifier determines its label.
Some of the main state-of-the-art detectors that use the region proposal approach are R-CNN~\cite{girshick2014rich}, Fast R-CNN~\cite{girshick2015fast}, and Faster R-CNN~\cite{ren2015faster}.
One-stage detectors do not use such a preprocessing stage; they instead use a single convolutional neural network (CNN) and directly output the prediction after a single pass of the input image.
A state-of-the-art one-stage detector is YOLO (\cite{redmon2016you, redmon2017yolo9000, redmon2018yolov3, bochkovskiy2020yolov4}).

YOLO (you only look once) is a state-of-the-art, real-time, one-stage object detector. 
It frames the object detection as a problem which predicts the bounding boxes and the class probabilities directly from a single pass on the input image. 
As a result, YOLO is both faster than other detectors and a real-time detector. 
YOLO divides the input image to a grid cell sized \textit{S}x\textit{S} in which each cell is responsible for detecting multiple objects. 
In YOLOv2~\cite{redmon2017yolo9000}, each cell predicts five options for detected objects.
Each option contains a vector that consists of the objectness score, which indicates the likelihood of the box containing an object, and the position and size of the box. 
In addition, each option contains a class probability vector that determines the class of the predicted object. 
This means that each cell can detect up to five different objects.
In the postprocessing stage, predictions are discarded if their objectness score is lower than a fixed threshold, and then the remaining boxes are fed to a NMS (non-maximal suppression) algorithm to choose the most appropriate predictions.
The rest are discarded and finally, the object detection output consists of the remaining predictions.
In this research we focus on attacking the YOLO object detection model.

\begin{figure}[h]
  \centering
\includegraphics[scale=0.8]{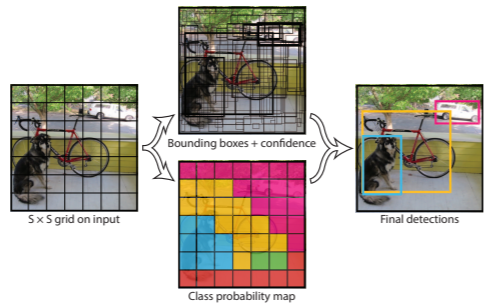}
    \caption{An illustration of YOLO object detector output (from \cite{redmon2016you}).}
    \label{fig:yolo}
\end{figure}

\subsection{\label{subsec_realWorldAttacks}Adversarial Attacks on Object Detectors}
Image classifier output is mainly a vector of the class probabilities. 
Hence, adversarial attacks on image classifiers usually use cross-entropy loss between the output and the target one-hot vector desired in the attack. 
In contrast, object detector output is more complicated.
As previously stated, it contains the bounding box positions, the object probability, and the class probability. 
A one-hot representation vector does not fit such output; therefore, adversarial attacks on object detectors use different approaches.

\citet{eykholt2018physical} extended the RP2 algorithm~\cite{eykholt2018robust}, which attacks image classification, to attack the YOLOv2 object detector in the real world.
They presented a real-world attack in which adversarial stickers are printed and attached to a stop sign. 
The authors also examined the transferability of the attack to the Faster R-CNN object detector.
Although~\cite{eykholt2018physical} achieved good results when tested in the real world, the adversarial stickers were not designed to be robust to extreme environmental noise, which includes different lighting conditions, camera noise, shadows, etc.
Hence, these attacks may fail when the stickers are tested under those conditions.
\citet{zhao2018seeing} considered those distortions and created a robust adversarial patch that attacks YOLOv3 in real-world settings. 
To achieve this, the attack used the EOT method~\cite{athalye2017synthesizing} and simulated different real-world distortions. 
The main target object in this study was also a stop sign.

While most of the real-world adversarial attacks against object detectors choose the stop sign as the attack target, 
\citet{thys2019fooling} developed an attack against the YOLOv2 detector in which people were targeted. 
The attack generates an adversarial patch that is printed, attached to the person, fed into the model, and causes the model to incorrectly detect or classify the person.
Its adaptation to the real world was also done using the EOT method~\cite{athalye2017synthesizing}.
The authors optimized the objective function using a dataset of different sized people in various poses, creating a patch that is robust to diverse instances of people, in contrast to an instance-specific attack, which is usually trained on a specific instance of an object.

We present a novel attack method that causes misdetection of large, nonplanar objects, in which dynamic adversarial patches are implemented by attaching screens to the target object.
Table \ref{tab:comparison} compares our attack to the attacks presented in other papers.

\begin{table*}[!ht]
\begin{tabularx}{\textwidth}{@{}l*{15}{C}c@{}}
\toprule
\diagbox[width=8em]{Paper}{Category} & 
{Attacks Image Classification}  & {Attacks Object Detection}  & {Nonplanar Objects}  &
{Real World}  & {Printer Constrained}  &
{Wide Range of Angles}  & {Dynamic} \\ \midrule
\citet{brown1712adversarial} & \checkmark &  &  & \checkmark & \checkmark &  & \\ 
\citet{athalye2017synthesizing} & \checkmark &  & \checkmark & \checkmark & \checkmark & \checkmark  &  \\ 
\citet{eykholt2018robust} & \checkmark &  &  &\checkmark & \checkmark & \checkmark & \\
\citet{eykholt2018physical} &  & \checkmark &  & \checkmark & \checkmark &  &  \\ 
\citet{thys2019fooling} & & \checkmark & \checkmark & \checkmark & \checkmark & \checkmark & \\ 
\citet{huang2020universal} & & \checkmark & \checkmark & \checkmark & \checkmark & \checkmark &  \\ 
\bottomrule
\textbf{Dynamic Adversarial Patch} & & \checkmark & \checkmark & \checkmark &  & \checkmark  & \checkmark\\ 
\bottomrule
\end{tabularx}
\caption{Comparison of related work.}
\label{tab:comparison}
\end{table*}

\section{\label{sec:threatmodel}Threat Model}
We assume an attacker that has full knowledge of the target (attacked) model, i.e., a white-box attack.

The attack can be performed in two main scenarios that differ in terms of the positioning and movement of the target object and the camera:
\begin{enumerate}
\item \textit{The target object is stationary, the camera is in motion:} in such a case, the camera's position changes over time, and hence the target object should identify it at a given moment and change the patches accordingly, in order to evade detection from the object detector that processes the camera's output. 
An example of this scenario is a parked car that is within view of a tracking drone.
\item \textit{The target object is in motion, the camera is stationary:} here, the target object (which wishes to evade detection) is aware of its changing position in relation to the camera and switches the patches when needed.
An example of this scenario is a truck traveling on a highway whose driver wants to dodge a stationary speed camera.
\end{enumerate}

Note that both the target object and the camera can be either stationary or in motion simultaneously.
Such cases resemble the above scenarios.
In order to use the dynamic patch, the target object should have a subsystem that can identify the camera's position or its own position relative to the camera; to simplify the experiment, we did not use such a sensor to determine the camera's position.
In this study, we focused on developing the attack, and thus we manually switched the patches according to the camera's position when filming (we mimicked the camera detection subsystem, as its development is beyond the scope of this paper).
In our experiments we focus on the first scenario; the attack is implemented in a similar way for the second scenario.

\section{\label{sec:attackMethod}Attack Method}

\subsection{Patch Generation Pipeline}
\begin{figure*}[h]
  \centering
\includegraphics[scale=0.47]{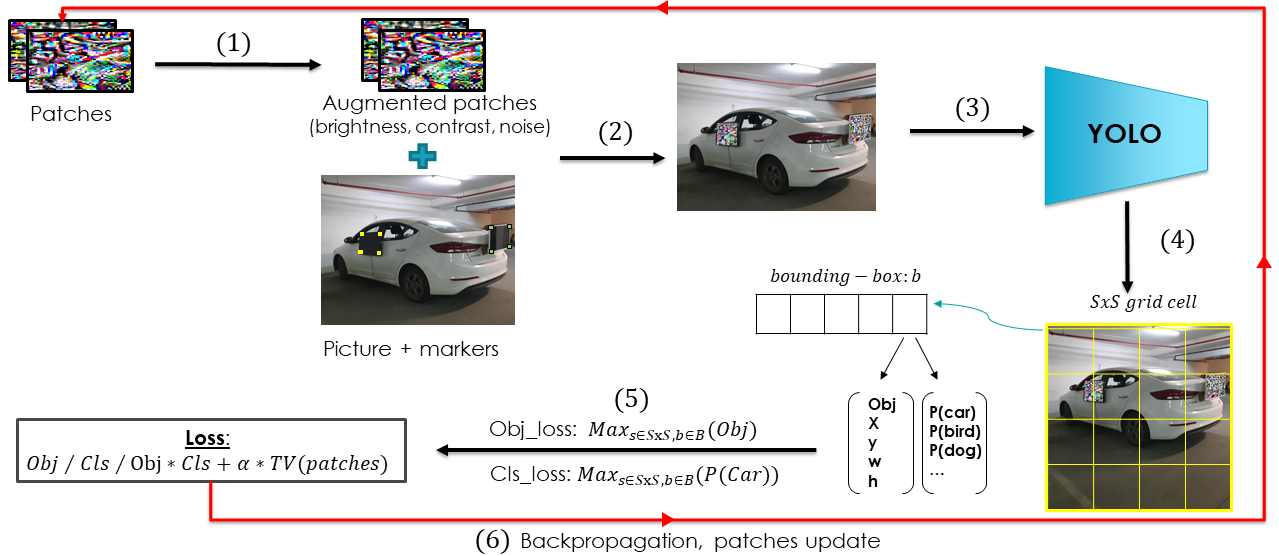}
  \caption{Patch generation pipeline.}
  \label{fig:pipeline}
\end{figure*}

The patch generation algorithm generates patches to be presented on each screen attached to the target object, optimized for a specific view angle.
This process is illustrated in Figure~\ref{fig:pipeline}. 
Given a dataset consists of images of the target object (with screen placeholders attached) viewed from a specific angle and the number of screens used, the process starts with generating random patches according to the number of screens. 
Then, a transformation function (presented in subsection~\ref{subsec_transormations}) is applied on the patches (step 1). 
The patches are carefully placed on the current batch of images, exactly on the flat screen placeholders, using a placing algorithm (presented in subsection~\ref{subsec_placingAlgo}) (step 2).
Next, the batch of patched images is sent to YOLO (step 3). 
The objective function (presented in subsection~\ref{subsec_objectiveFunc}) is computed with the model's output serving as the input for the loss function (steps 4 + 5). 
Using the optimization algorithm over the whole (differentiable) pipeline, the patch's pixels are changed accordingly (step 6). 
For a formal description of the generation pipeline, see Algorithm~\ref{alg:PatchGeneration} in the Appendix.

\subsection{\label{subsec_transormations}Using Transformations to Make the Attack More Robust}
Real-world settings often include different factors that influence the captured pictures - from environment conditions like sunlight and shadows, to device limitations like camera noise.
This is exacerbated when presenting the patches on a screen instead of printing them.
The added difficulty results in noise on the screen originating from the camera's reflection or sunlight, which adds brightness to the screen.
To simulate real-world settings in the training process, we apply a transformation function to the patch in each training iteration. The function randomly changes the brightness and contrast of the patch, in addition to adding random noise to it.

\subsection{\label{subsec_placingAlgo}The Placing Algorithm}
Through the patch generation pipeline, patches are placed on images in the training set.
To achieve maximum resemblance of the patched training images to the real-world setting (projection of patches on screens attached to a car), the patches should be attached accurately to the car image, precisely within the borders of the screens.
In our patch generation process, we use the placing algorithm used in \cite{kaziakhmedov2019real} to obtain accurate results that simulate the real-world situation.
For each image in the training set, we manually mark the four edges of each screen by changing the color of a single pixel on each screen's edge to a fixed color.
This marks the borders of the screens.
The marked image and the patches are fed to the algorithm which places the patches exactly in the position of the screens (using image masking and perspective transformation of the patch images).
Note that the algorithm, as a part of the patch generation pipeline, is differentiable, to enable the patches optimization.

\subsection{\label{subsec_objectiveFunc}Objective function}
To carry out the attack, we define an objective function, which is minimized during the training phase; this process optimizes the patch.
The objective function consists of a few components:\\
\subsubsection{\textbf{The Loss Function}}
Eykholt et al. \cite{eykholt2018physical} extended the RP-2 algorithm~\cite{eykholt2018robust} to enable its use for attacking object detectors. 
As previously mentioned, YOLO object detector divides the input image into  \textit{S}x\textit{S} grid cells, where each cell contains predictions for \textit{B} bounding boxes, and each bounding box contains the objectness score, the position of the box, and a class probability vector. 
The attack used in~\cite{eykholt2018physical} minimizes the maximum class probability of the target object (which wishes to evade detection) over the predictions in the whole image, which causes YOLO to associate the predicted object incorrectly to the wrong class or even fail to detect the object at all.
They formulate this loss function as:
\begin{equation}
J = max_{s\in SxS,  b \in B}P(s, b, y, f_\theta(x))
\label{eq:eykholtLoss}
\end{equation}
where \textit{x} is the perturbed image, \textit{y} is the target object's class, \textit{f} is the model function, and \textit{P} is the function that extracts the class score.

Thys et al. \cite{thys2019fooling} extended this attack by presenting three different options for the loss function:
\begin{enumerate}
\item \textbf{Class loss \textit{(cls)}} - as described in Equation \eqref{eq:eykholtLoss}.
\item \textbf{Object loss \textit{(obj)}} - outputs the maximum objectness score over the predictions in  the whole image. 
By minimizing that score for a bounding box prediction \textit{B} below the model's detection threshold, the model will consider this box empty of objects, causing the model to ignore it. 
Extracting the maximum value of this score for all of the bounding boxes \textit{B} in each cell \textit{S} will cause the object to ``disappear" and avoid detection.
\item \textbf{Object-Class loss \textit{(obj\_cls)}} - outputs a combination of the two above losses. 
The combination chosen is the product of the losses.
\end{enumerate}
In our attack, we let the loss function be one of these options and examine the attack performance regarding each loss.\\
\subsubsection{\textbf{Total Variation}}
The success of this attack in the real world necessitates that the camera captures (with reasonable quality) the patch. 
To make it easier for the camera, we want to make the patches smooth and contain ``regions" consisting of pixels of similar colors. 
This way the patch can be captured clearly, as opposed to a noisy patch that contains sharp color transitions and many different colored pixels in close proximity to one another, which won't be captured clearly.
To do so, we use the total variation~\cite{mahendran2015understanding} loss for patch \textit{p}:

\begin{equation}
TV(p) = \sum_{i, j }\sqrt{(p_{i,  j} - p_{i+1,  j})^2 + (p_{i,  j} - p_{i,  j+1})^2}
\label{eq:TV}
\end{equation}
Minimizing the total variation of a patch means reducing the distance between every two nearby pixels' colors in the image, causing the effect of smoothness.\\

{\textbf{No Printer Limitation}}
Most of the existing attacks use printed noise as the perturbation in the real world \cite{eykholt2018physical,thys2019fooling, eykholt2018robust,sharif2016accessorize}.
The colors that most printers can produce are only a subset of the whole RGB color space.
Thus, those attacks use the NPS (non-printability score) \cite{sharif2016accessorize} as part of the objective function to deal with the above limitation.
Considering that we use a screen to present the patches, we didn't find it necessary to use the NPS or another similar score in the attack.\\

To sum up, our objective function for patch \(p\), patched input image \(x_{p}\), and model function \(f\) with weights \(\theta\) is:
\begin{equation}
J(p, x_{p}) = \alpha * TV(p) + Loss(f_{\theta}(x_{p}))
\label{eq:OurobjectiveFunction}
\end{equation}
where \(Loss\) stands for \textit{obj} loss, \textit{cls} loss, or \textit{obj\_cls} loss.
\(\alpha\) is determined empirically.
We optimize the objective function using the Adam optimizer \cite{kingma2014adam}.

\subsection{Semantic Adversarial Patch}
During the evaluation we noticed that in some video frames the \textit{car} object was classified as a \textit{bus} or \textit{truck}.
In terms of the attack's success, those frames are regarded as successful frames, as the \textit{car} is misclassified.
However, \textit{bus} and \textit{truck} are semantically-related to the original class, \textit{car}.
Such attack has no significant negative influence on the attacked system's behavior in some cases, such as when an autonomous car detects a \textit{truck} instead of a car.
Therefore, we extended our objective function to specifically create patches that prevent the model from classifying the target object as any semantically-related object.
This was done by adding a component which is responsible for minimizing the detection of each of the semantically-related classes:

\begin{equation}
J(p, x_{p}, C) = \alpha * TV(p) + \sum_{y_i\in C }Loss(f_{\theta}(x_{p}), y_i)
\label{eq:semanticLossFunction}
\end{equation}
where \(C\) is the set of labels we want to prevent the model from using during the classification of the target object.
The equation is similar to \eqref{eq:OurobjectiveFunction}.
This way, eventually a car will be classified, for example, as a cake or person instead of a car or a semantically similar object such as a bus or truck.

\section{Dynamic Adversarial Patch}
Previously proposed real world adversarial attacks on object detectors share two main properties that significantly narrow the scope of the attack:
\begin{enumerate}
\item \textbf{The attacked objects} - In \cite{eykholt2018physical} and \cite{zhao2018seeing}, the main target object is a stop sign.
While their goal is similar to ours: making a detector fail to identify an existing object in a real-world scene (i.e., a hiding attack), most of their attacked objects are planar (commonly a stop sign).
\item \textbf{The perturbation} - For the perturbation, \cite{eykholt2018physical} and \cite{zhao2018seeing} use adversarial stickers, \cite{zhao2018seeing} uses also an adversarial poster, and \cite{thys2019fooling} and \cite{brown1712adversarial} use an adversarial patch.
They all use printed, fixed-position, and static adversarial noise (i.e., the content of the patch is fixed).
\end{enumerate}

The proposed \textbf{dynamic adversarial patch} is aimed at both planar and nonplanar objects, using screens to present the pregenerated patches and switch between them in accordance with the camera view angle.
By attaching screens to different planes of the object, we make the object invisible to detectors from different views.
By switching the patches during the attack according to the object detection system's position, the attack presents the optimal combination of patches and adapts itself \textit{dynamically} to the current situation.

The patches are created through a process of splitting the training set into subsets and computing patches for each subset, where each subset represents a different view angle of the camera.
Our patch generation algorithm (Algorithm~\ref{alg:PatchGeneration}) generates patches for each subset (according to the number of screens). 
Then, the attack (using the generated patches) is evaluated over a test set that has been divided into subsets accordingly.
If the split is ineffective (i.e., achieves a lower success rate than the previous split), the algorithm stops and returns the previous split's patches (and by doing so, determines the number of times the patches should be switched).
Otherwise, it splits the training set into more subsets and trains patches accordingly.
The formal algorithm is described in Algorithm \ref{alg:dynamicAttack}.

\begin{figure}[h]
  \centering
\includegraphics[scale=0.4]{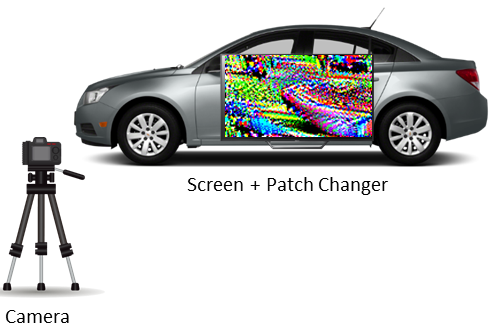}
    \caption{An illustration of the testbed.}
    \label{fig:testbed}
\end{figure}

\begin{figure}[h]
  \centering
  \begin{subfigure}[b]{0.24\textwidth}
    \centering
\includegraphics[width=\textwidth]{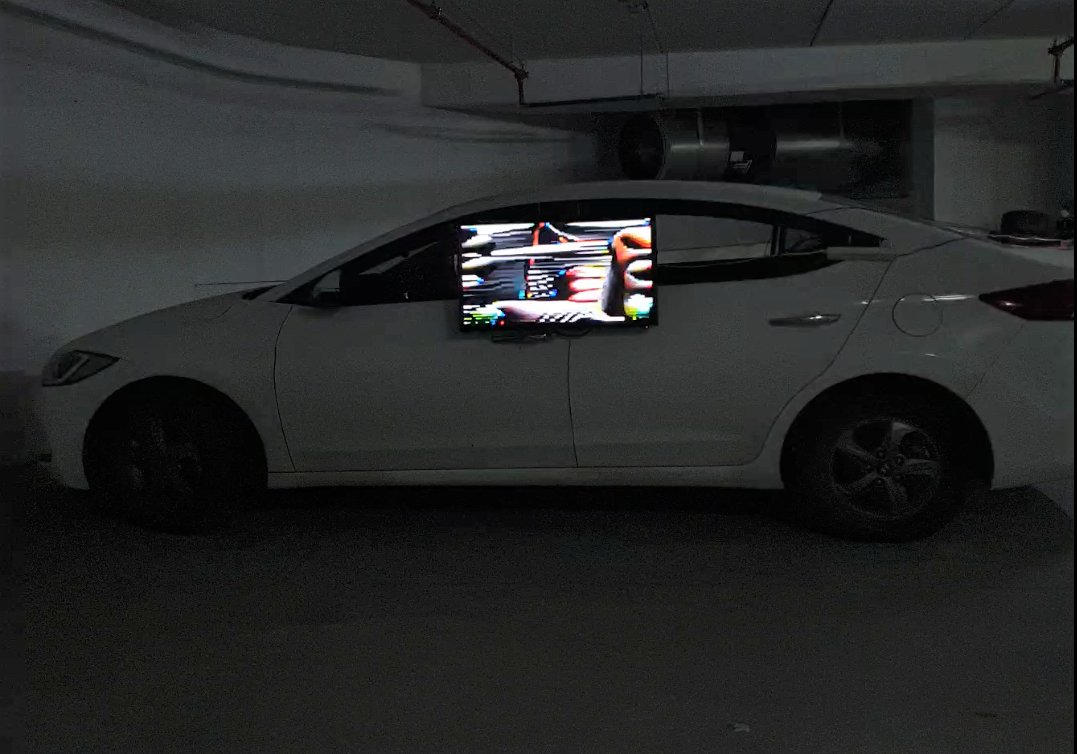}
    \caption{the left side of the car}
  \end{subfigure}
  \begin{subfigure}[b]{0.24\textwidth}
    \centering
    \includegraphics[width=\textwidth]{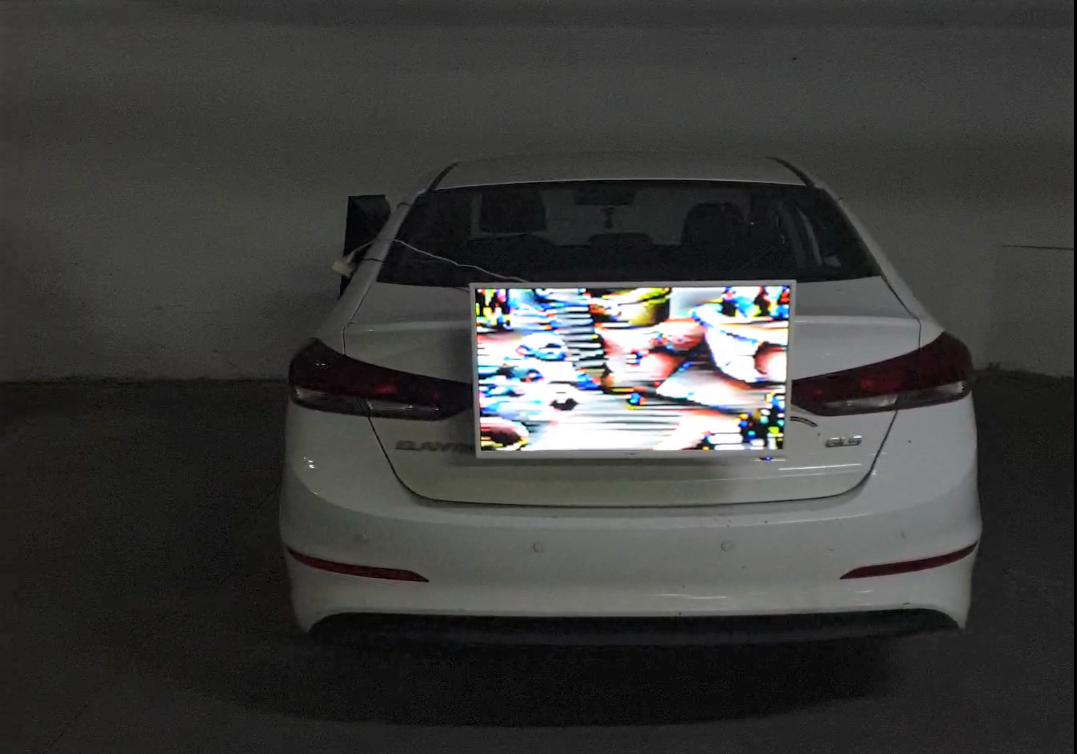}
    \caption{the back of the car}
  \end{subfigure}
  \caption{The testbed in the real world.}
  \label{fig:realworldtestbed}
\end{figure}

\section{\label{sec_eval}Evaluation}
We evaluate our attack using a car in an underground parking lot.
There are different options for attaching the screens to the car.
In the course of our experiments, we opted to attach screens on the car's sides and on the car's back.
To evaluate the attack, we took videos of the car while moving the camera around it, maintaining a fixed distance of four meters from the target car.
We evaluated our attack using YOLOv2, which was trained on the MS COCO dataset~\cite{lin2014microsoft}. 
The detection threshold was set at 0.5.
\\ \\
\textbf{Metrics.} We measured the attack performance by calculating the attack success rate for a set of video frames - the percentage of frames (images) in which the target object (car) was not detected as the true class (car). 
Let \(VF\) be the set of video frames of a video containing the target object. 
For model \(f\) with parameters \(\theta\), the attack success rate is:
\begin{equation}
\frac{|\{vf \in VF \mid car \notin f_{\theta}(vf)\}|}{|VF|} * 100
\label{eq:attackSuccessRate}
\end{equation}
\subsection{Experimental Setup}
\textbf{Testbed.} To build the training and testing sets and perform the evaluation, we built a testbed, which is illustrated in Figure~\ref{fig:testbed}; Figure \ref{fig:realworldtestbed} presents the testbed in the real world.
The testbed contains the following:
\begin{itemize}
\item Cars: Hyundai Elantra 2016/Toyota Corolla 2006/Fiat Tipo 2018.
\item Screens: MASIMO LE32FD and Jetpoint JTV2828, respectively sized 32" and 28".
\item Patch Changer: a mini PC that displays and switches between the various pre-computed adversarial patches.
\item Camera: a standard Samsung Galaxy Note 9 camera.
\end{itemize}
The screens were attached to the car and connected to the Patch Changer.
\\ \\
\textbf{Dataset preparation.} We used two different datasets - one for training and one for testing, each consists of data within video frames:
\begin{itemize}
    \item 
        Training Dataset: Our method of presenting patches on a screen necessitated that we prepare a training dataset for a car with screens attached from various angles.
        As previously explained, the images in the training dataset are manually annotated (with four pixels that mark the borders of the screen) to enable the placing algorithm to accurately place the patch on the images.
        Preparation of the training dataset included placing screens on the chosen locations on the car and taking videos of it.
        We filmed around the car in order to prepare optimized patches from many angles.
        The video was then cut into frames (we cut it to 20 FPS - 471 frames), and as explained above, we manually marked the four borders of the screens in each image.
        These images form the training dataset.
    \item
        Testing Dataset: The testing set was prepared similarly to the training set. 
        We took videos of the car, which had screens attached to it, while presenting the appropriate patches on the screens.
        The video was then cut into frames; the evaluation was performed on these frames.
        \\
\end{itemize}
\textbf{Types of Experiments Performed.}
\begin{itemize}
\item \textit{Digital world} - These experiments helped us evaluate our attack before applying it in the real world.
The test images were cut from a video in which the car was presented with placeholders (i.e., the screens were attached but no patches were presented on them).
In these experiments, the patches generated were attached to the test set images (using the placing algorithm) in the exact spot where the screens are located.
Then, we evaluated the attack, relating to each image as a single frame of a video.
\item \textit{Real world} - In order to evaluate our work, we applied the attack in the real world by placing screens on a car, presenting the optimized patches, and changing the patch on each screen, depending on the view angle during the video filming.
The videos were then cut into frames and evaluated with the object detection model.
Then, the success rate was calculated by \eqref{eq:attackSuccessRate}.
\end{itemize}

Note that we conducted the experiments in an underground parking lot, in which we cannot control the environmental conditions, such as lighting and other environmental noise.
However, due to the lack of sunlight, the environmental conditions in the underground parking lot simulate a cloudy day without strong sunlight. 
\begin{figure}[h]
  \centering
  \begin{subfigure}[b]{0.18\textwidth}
    \centering
\includegraphics[width=\textwidth]{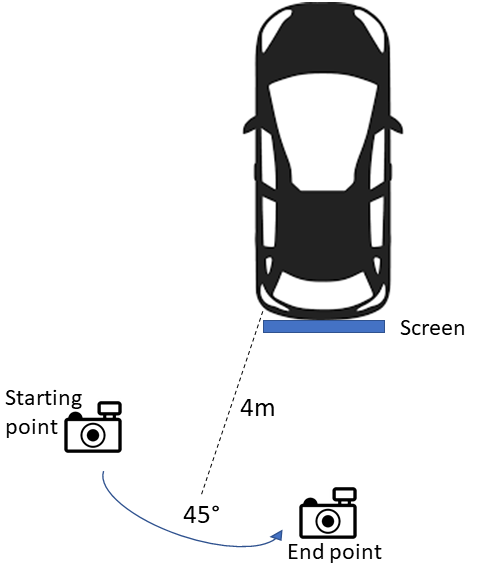}
    \caption{one screen setting}
    \label{fig:oneScreenSubFig}
  \end{subfigure}
  \begin{subfigure}[b]{0.18\textwidth}
    \centering
    \includegraphics[width=\textwidth]{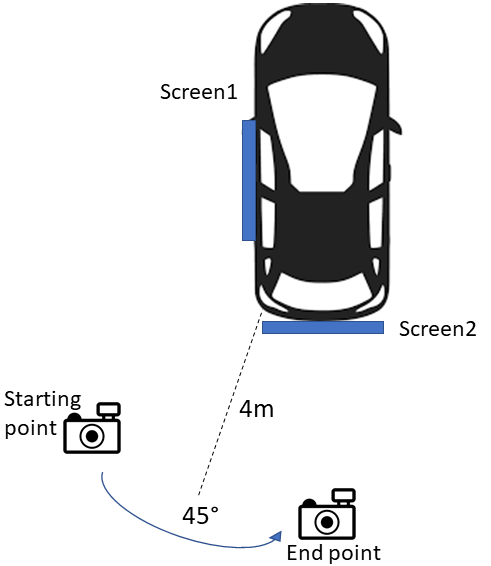}
    \caption{two screen setting}
    \label{fig:TwoScreensSubFig}
  \end{subfigure}
  \caption{Experiment evaluating the impact of the number of screens.}\label{fig:NumOfScreensExp}
\end{figure}

\subsection{The Experiments}
We test our attack with a few different configurations to find the optimal one for carrying out a successful attack:
\\ \\
\textbf{Number of screens.} We checked whether one screen is sufficient to achieve good results when filming from various angles or whether a combination of more than one screen is needed.
We examined this issue by evaluating the attack as follows: targeting a parked car with a screen attached to its back and a parked car with two screens attached, one on its back and one on the car's left side.
The two settings are illustrated respectively in Figure~\ref{fig:oneScreenSubFig} and Figure~\ref{fig:TwoScreensSubFig}.
The patches used in this experiment were not dynamic, i.e, the screens presented only a single patch throughout the video, without switching, to resemble the adversarial patch method used in many recent papers (described in Section~\ref{sec_relatedWork}).

The results are presented in Table~\ref{table:numOfScreens}.
As can be seen in the table, with \textit{obj\_cls} loss, the car with one screen attached obtained a success rate of 43.4\%.
The car with two screens attached achieved a higher success rate of \textbf{74\%}.
For the other loss functions, the attack with two screens also obtained higher success rates.
These results confirm that using a single screen may not be effective when attacking large, nonplanar objects from a wide view range.
\begin{table}[h!]
\centering
\begin{tabular}{|c | c | c | c |} 
 \hline
 & \textit{obj} loss & \textit{cls} loss & \textit{obj\_cls} loss \\ [0.5ex] 
 \hline\hline
 one screen  & 24.8\% & 37\% & \textbf{43.42}\% \\
 \hline
 two screens & 28.45\% & 62.39\% & \textbf{74}\% \\ 
 \hline
\end{tabular}
\caption{Attack success rate for different numbers of attached screens and various loss functions.}
\label{table:numOfScreens}
\end{table}
\begin{figure}[h]
  \centering
  \begin{subfigure}[b]{0.23\textwidth}
    \centering
\includegraphics[width=\textwidth]{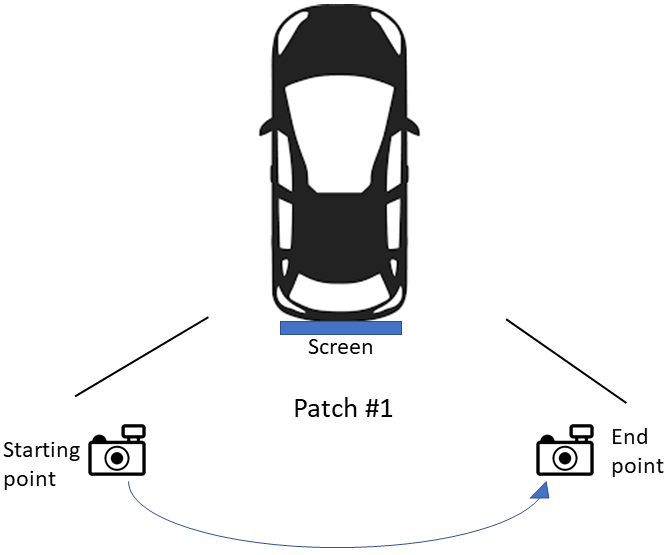}
    \caption{one patch}
    \label{fig:numOfPatches1}
  \end{subfigure}
  \begin{subfigure}[b]{0.23\textwidth}
    \centering
    \includegraphics[width=\textwidth]{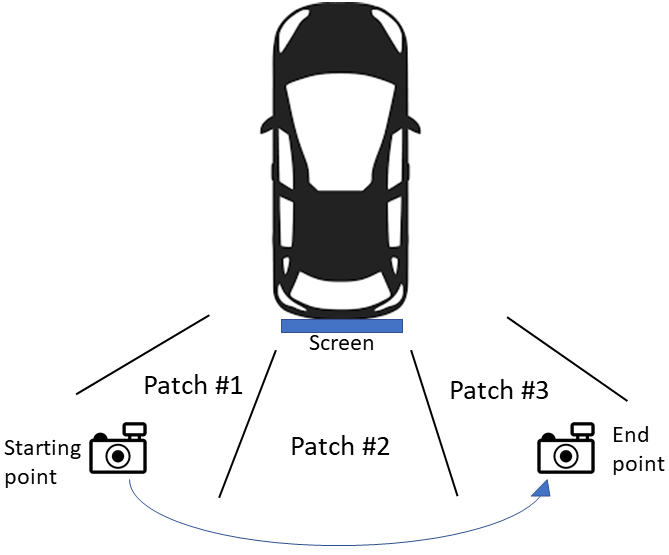}
    \caption{three patches}
    \label{fig:numOfPatches3}
  \end{subfigure}
  \begin{subfigure}[b]{0.23\textwidth}
    \centering
    \includegraphics[width=\textwidth]{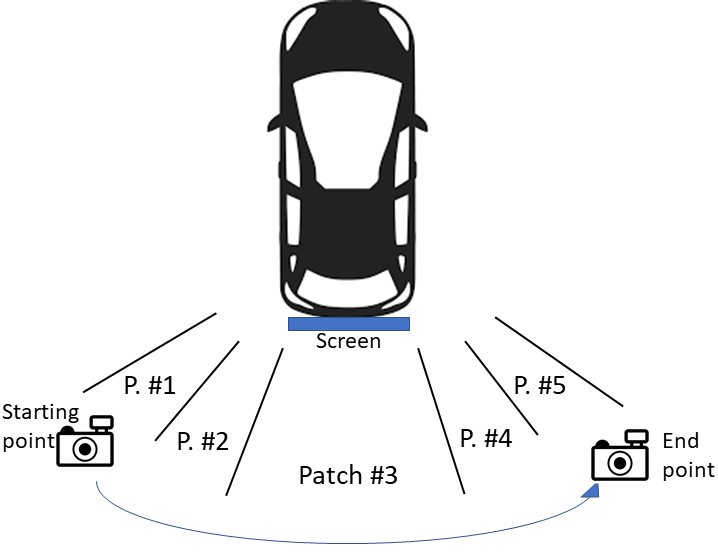}
    \caption{five patches}
    \label{fig:numOfPatches5}
  \end{subfigure}
  \begin{subfigure}[b]{0.23\textwidth}
    \centering
    \includegraphics[width=\textwidth]{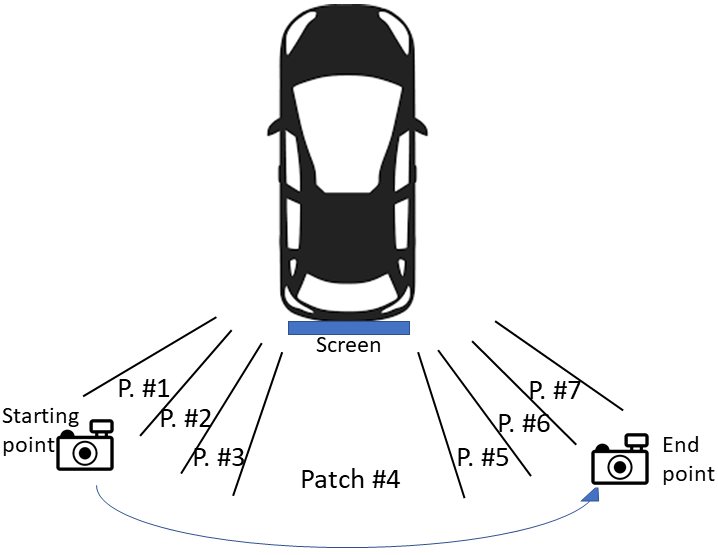}
    \caption{seven patches}
    \label{fig:numOfPatches7}
  \end{subfigure}
  \caption{Experiment evaluating the impact of the number of patches.}\label{fig:numOfPatchesExp}
\end{figure}
\\
\textbf{Number of patches.} In this experiment, the ``dynamic" aspect of the attack is analyzed.
We want to assess whether switching between patches optimized for a specific view angle achieves better results than presenting a single patch, and determine the optimal number of patches to present.
To do so, we first train a single patch over the entire view angle range.
Then, we divide the training set into two subsets, train a patch for each set and evaluate the attack using those patches.
We do this repeatedly, each time the training set is divided into additional subsets.
We apply the attack both in the digital space and in the real world. 
The experiment is illustrated in Figure~\ref{fig:numOfPatchesExp}.
\begin{figure}[h!]
  \centering
\includegraphics[scale=0.7]{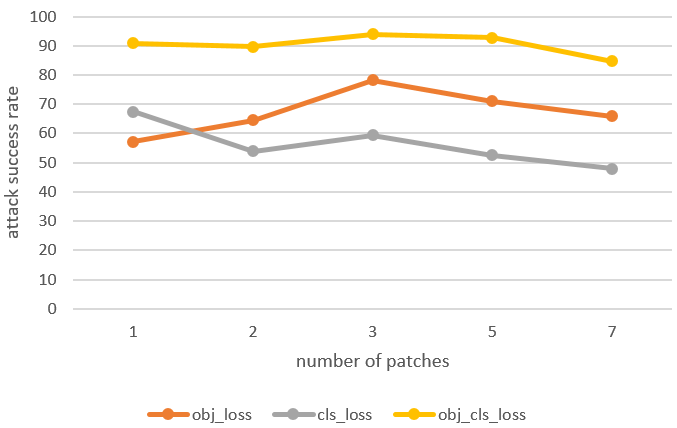}
  \caption{The impact of the number of patches on the attack success rate for various loss functions (in the digital space).}
  \label{fig:numOfPatchesDigitalResults}
\end{figure}

Figure~\ref{fig:numOfPatchesDigitalResults} presents the digital space results.
Indeed, most of the times, the patch switching was effective.
We can also see that over-splitting the view range dataset (to more than 3 subsets) decreases the attack success rate.

\begin{figure}[h]
  \centering
  \begin{subfigure}[b]{0.21\textwidth}
    \centering
\includegraphics[width=\textwidth]{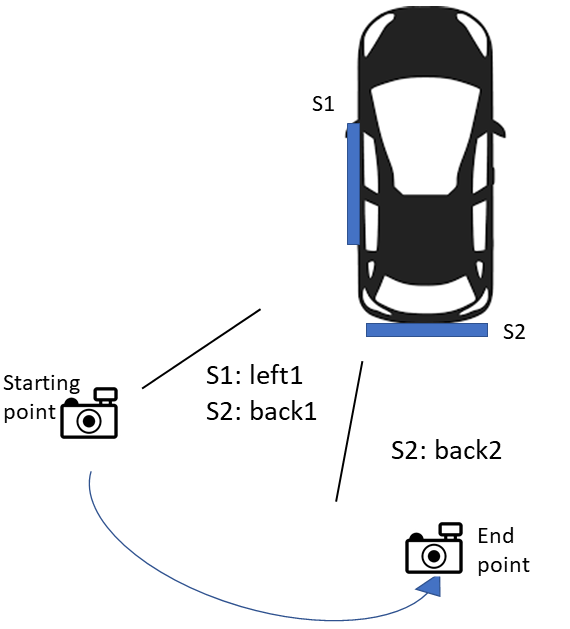}
    \caption{two patches}
    \label{fig:numOfPatchesTwoScreens1}
  \end{subfigure}
  \begin{subfigure}[b]{0.21\textwidth}
    \centering
    \includegraphics[width=\textwidth]{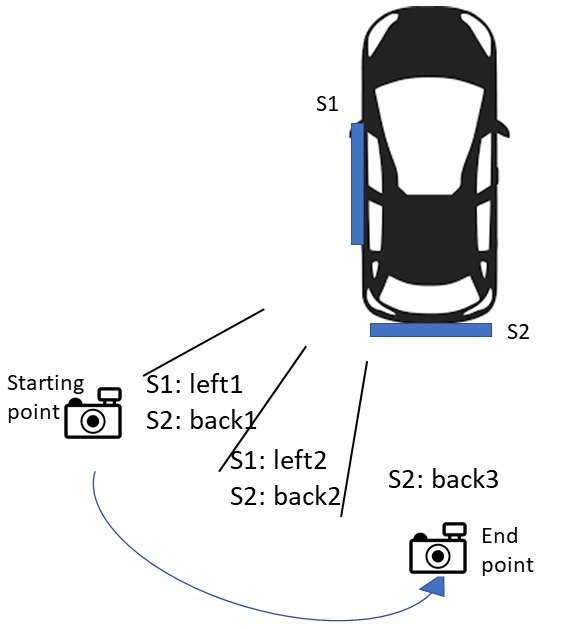}
    \caption{three patches}
    \label{fig:numOfPatchesTwoScreens2}
  \end{subfigure}
  \caption{Experiment evaluating the impact of the number of patches (with two screens).}\label{fig:2ScreensNumOfPatchesExp}
\end{figure}

\begin{figure}[h]
  \centering
\includegraphics[scale=0.23]{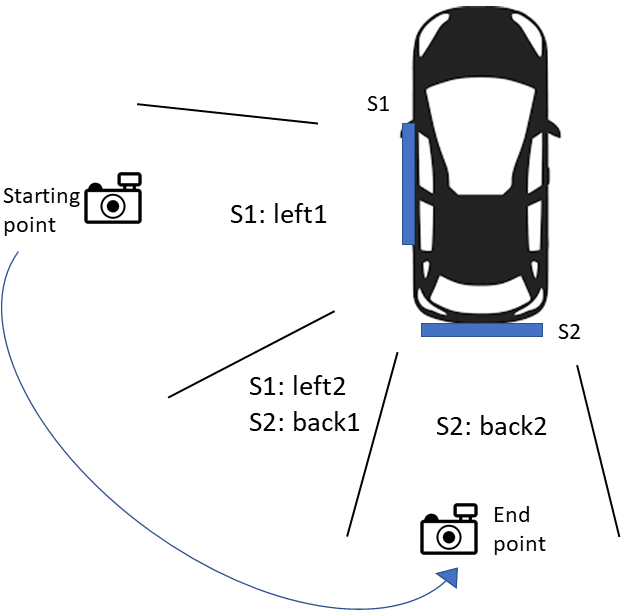}
  \caption{90\degree Attack.}
  \label{fig:fullRangeAttack}
\end{figure}

When we applied the attack in the real world (i.e., presented the patches on the screen), the results were unsatisfactory.
We observed that the attack almost always failed at the wider angles on both the left and right sides of the car, when they became visible during filming.
In most cases, the attack was successful when the camera was in the middle of its route, when most of the back of the car was visible.
\\\\
\textbf{Wide-Range Real World Attack.}
Combining the conclusions from the two former experiments, we conclude that in order to successfully perform the attack in the real world from a wide range of view angles, more than one screen is needed and more than one patch is required.
Thus, we perform the following experiments:
\begin{figure}[h!]
  \centering
  \begin{subfigure}[b]{0.23\textwidth}
    \centering
\includegraphics[width=\textwidth]{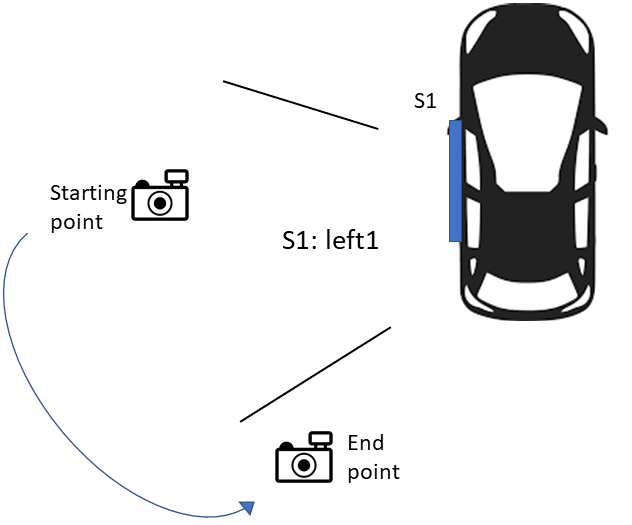}
    \caption{no replacement}
    \label{fig:NumOfPatchesSideExpNoRep}
  \end{subfigure}
  \begin{subfigure}[b]{0.23\textwidth}
    \centering
    \includegraphics[width=\textwidth]{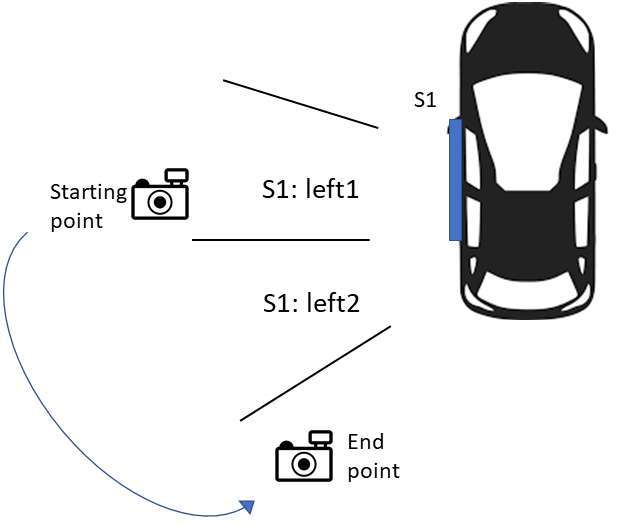}
    \caption{one replacement}
    \label{fig:NumOfPatchesSideExpOneRep}
  \end{subfigure}
  \caption{Experiment evaluating the impact of the number of patches - side view experiment.}
  \label{fig:NumOfPatchesSideExp}
\end{figure}
\\\\
\textbf{Attack From the Left.} This experiment determines how many patch switches are required for a successful attack from a view range of the car's side.
Figure \ref{fig:NumOfPatchesSideExp} describes the experiment, and Table \ref{table:leftRange_results} presents the results.
We observe that in this case, to successfully attack from the side, no patch switching is required.
\begin{table}[h!]
\centering
\begin{tabular}{|c | c | c | c |} 
 \hline
 \# of patches & \textit{obj} loss & \textit{cls} loss & \textit{obj\_cls} loss \\ [0.5ex] 
 \hline\hline
 1 & 0\% & 9.5\% & \textbf{90}\% \\ 
 \hline
 2 & 0\% & 5\% & \textbf{12\%} \\
 \hline
\end{tabular}
\caption{Attack success rate with different losses, left side.}
\label{table:leftRange_results}
\end{table}
\\
\textbf{Attack From the Back.} This experiment determines how many patch switches are required for a successful attack from a view range of the car's back (45\degree).
Figure \ref{fig:2ScreensNumOfPatchesExp} describes the experiment, and Table \ref{table:2screens45Deg_results} presents the results.
In this view range, the results approve that switching between two patches indeed improves the results.
\begin{table}[h!]
\centering
\begin{tabular}{|c | c | c | c |} 
 \hline
 \# of patches & \textit{obj} loss & \textit{cls} loss & \textit{obj\_cls} loss \\ [0.5ex] 
 \hline\hline
 1 & 28.45\% & 62.39\% & \textbf{74}\% \\ 
 \hline
 2 & 40\% & \textbf{81\%} & 67.61\% \\ 
 \hline
 3 & 29.40\% & 35.35\% & \textbf{63.74\%} \\
 \hline
 4 & 25.13\% & 17.40\% & \textbf{51.89\%} \\
 \hline
\end{tabular}
\caption{Attack success rate with different losses, car's back.}
\label{table:2screens45Deg_results}
\end{table}
\\
\textbf{90\degree Attack.} Finally, combining the two former experiments, we created an attack that works for a full quarter of a circle, covering $90\degree$.
The attack is illustrated in Figure \ref{fig:fullRangeAttack}.
The patches themselves (and the number of switches) in each of the two sub-ranges are the ones that obtain the best results in each of the last experiments.
The results are presented in Table \ref{table:fullRangeCoverageResults}.
The best success rate for the last scenario is \textbf{80}\%, a satisfactory result when taking into account the wide range of the attack.
\begin{table}[h!]
\centering
\begin{tabular}{|c|c|c|c|} 
\cline{2-4}
    \multicolumn{1}{c|}{} & \textit{obj} loss & \textit{cls} loss & \textit{obj\_cls} loss \\ [0.5ex] 
 \hline 
    Success Rate & 40\% & 41.3\% & \textbf{80}\% \\ 
 \hline
\end{tabular}
\caption{Attack success rate with different losses, full range attack ($90\degree$).}
\label{table:fullRangeCoverageResults}
\end{table}
\\
\textbf{Most Successful Loss.} In all of the previous experiments, we considered the three types of loss functions: \textit{obj}, \textit{cls}, and \textit{obj\_cls}.
Most of the time, patches that were trained by the \textit{obj\_cls} loss seem to outperform patches trained by other loss types.
However, there isn't a specific answer to the question of which loss is better, as the \textit{cls} loss patches also gained some good results in the previous experiments.
In addition, to make the attack robust to a wide view range, patches from different loss types are combined.
For example, Figure \ref{fig:fullRangeAttack} covers the greatest attack range and consists of a few patches trained by different loss functions.
\\\\
\textbf{Semantic Adversarial Patch.}
We also examined the success rate of the \textit{semantic patches} that were computed so that the detector would classify the object's class to a semantically-unrelated class.
In our experiments, those patches are targeted at preventing the object detection model from classifying the car as any vehicle (\textit{car}, \textit{bus}, or \textit{truck}).
To evaluate the success of this kind of patch, we define a success rate metric which is derived from the attack success rate defined by Equation~\eqref{eq:attackSuccessRate}, in which a successfully attacked frame is a frame in which there is no detection of \textit{car}.
However, in this case, we defined a successfully attacked frame as a frame in which there is no detection of any of the following vehicles: \textit{car}, \textit{bus}, or \textit{truck}.
Therefore, the success rate is defined as the percentage of frames that do not contain any detection of \textit{car}, \textit{bus}, or \textit{truck}.

We performed the $90\degree$ view range experiment (Figure \ref{fig:fullRangeAttack}) while presenting the semantic patches.
From the results presented in Table \ref{table:semanticPatchResults}, we can observe that the semantic patches were effective as most of the classifications in the experiments were \textit{cake}, \textit{boat}, \textit{person}, and \textit{traffic light}.
\begin{table}[h!]
\centering
\begin{tabular}{|c|c|c|c|} 
\cline{2-4}
    \multicolumn{1}{c|}{} & \textit{obj} loss & \textit{cls} loss & \textit{obj\_cls} loss \\ [0.5ex] 
 \hline 
    Success Rate & 44.28\% & 35.17\% & \textbf{72}\% \\ 
 \hline
\end{tabular}
\caption{Semantic adversarial patch results, a full quarter of a circle ($90\degree$) coverage.}
\label{table:semanticPatchResults}
\end{table}

\begin{figure*}[h]
\begin{center}
\begin{subfigure}[b]{0.45\textwidth}
    \centering
\includegraphics[width=\textwidth]{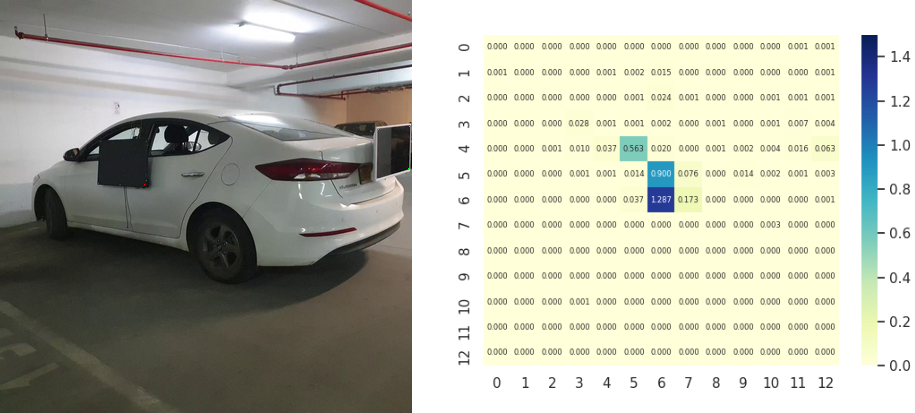}
    \caption{Heatmap of non-patched image}
    \label{fig:nonPatchedHeatmap}
\end{subfigure}
\begin{subfigure}[b]{0.45\textwidth}
    \centering
\includegraphics[width=\textwidth]{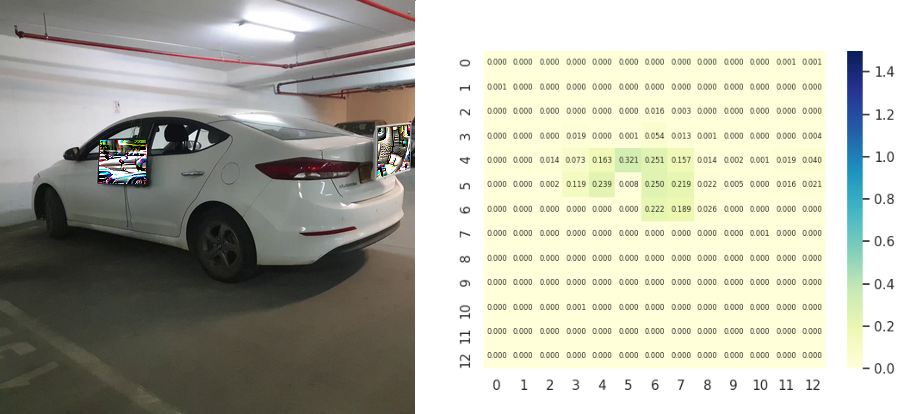}
    \caption{Heatmap of image with \textit{obj} loss patches}
    \label{fig:objHeatmap}
\end{subfigure}
\begin{subfigure}[b]{0.45\textwidth}
    \centering
\includegraphics[width=\textwidth]{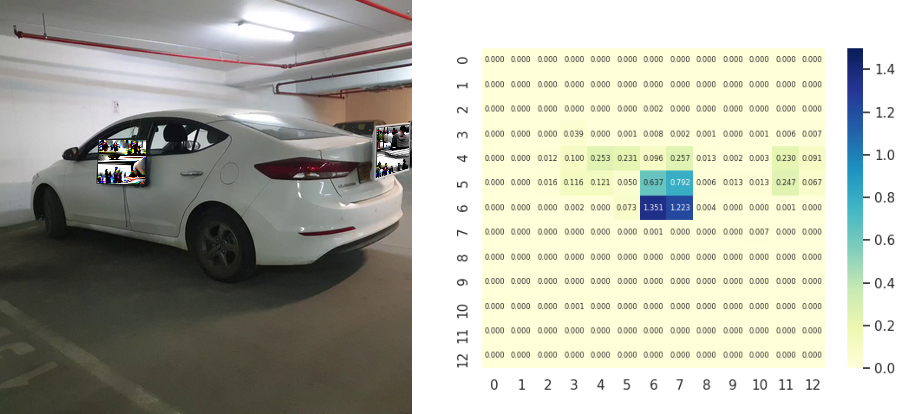}
    \caption{Heatmap of image with \textit{cls} loss patches}
    \label{fig:clsHeatmap}
\end{subfigure}
\begin{subfigure}[b]{0.45\textwidth}
    \centering
\includegraphics[width=\textwidth]{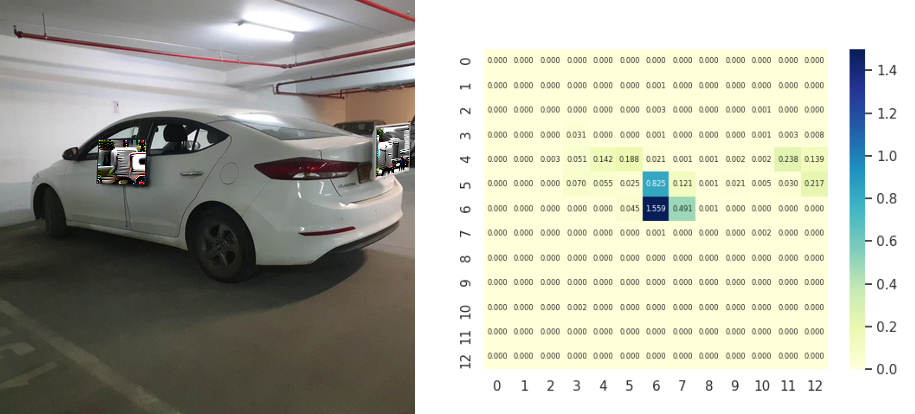}
   \caption{Heatmap of image with \textit{obj\_cls} loss patches}
   \label{fig:obj_clsHeatmap}
\end{subfigure}
\end{center}
\caption{The impact of different types of losses on YOLO's output.}
\end{figure*}

\section{Further Analysis}
\subsection{Attack Impact}
As described in section \ref{sec:attackMethod}, three types of loss functions can be used in the attack: \textit{obj} loss, \textit{cls} loss, and \textit{obj\_cls} loss, each of which has a different impact on YOLO.
While \textit{obj} loss is targeted at minimizing the objectness score over the whole image and hence at causing YOLO to misdetect any object, \textit{cls} loss is aimed at reducing the class score of a specific class in every cell in YOLO's grid.
Moreover, the goal of \textit{obj\_cls} loss is to minimize a combination of the objectness score and the class score of a target class in that bounding box, over the whole image. \par
To gain a deeper understanding of the effect of the attack, we performed an analysis of the impact of each loss on YOLO's output.
Recall that the output goes through a postprocessing stage in which some of the bounding boxes are filtered out due to a detection threshold that is higher than their objectness score, and then the most suitable bounding boxes are chosen for the prediction, using an NMS algorithm. \par
Our attack harms the output of YOLO itself, in such a way that each loss function type harms a different aspect of the output and may affect differently on the postprocessing stage (however, ultimately, the target object will not be detected or will be wrongly classified).
In order to perform an analysis of the impact of the three loss functions examined, we expressed YOLO's output as a heat map and examined the results.
Each cell in the map represents the appropriate cell from the grid YOLO creates.
The number inside each cell is the sum of the objectness scores from each bounding box that YOLO predicted for that cell. \par
First, we fed YOLO an image of a car without patches.
Then, we applied two patches digitally on the image, exactly within the screens' borders, using the placing algorithm (subsection \ref{subsec_placingAlgo}).
This was done three times - each time with the patches calculated using a different loss function, and we generated a heat map for each of the images.
The results are illustrated in Figures \ref{fig:nonPatchedHeatmap}, \ref{fig:objHeatmap}, \ref{fig:clsHeatmap} and \ref{fig:obj_clsHeatmap}.
In the figures it can be seen that, as expected, the \textit{obj} loss patches reduce the objectness score significantly over all of the cells.
As a result, no objects were detected at all.
\begin{figure}[h]
  \centering
  \begin{subfigure}[b]{0.22\textwidth}
    \centering
\includegraphics[width=\linewidth]{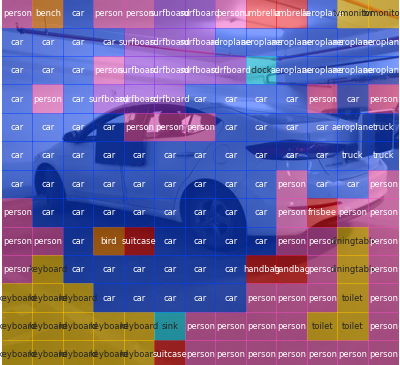}
    \caption{non-patched image}
    \label{fig:classes_heatmap_NO_patches}
  \end{subfigure}
  \begin{subfigure}[b]{0.22\textwidth}
    \centering
    \includegraphics[width=\linewidth]{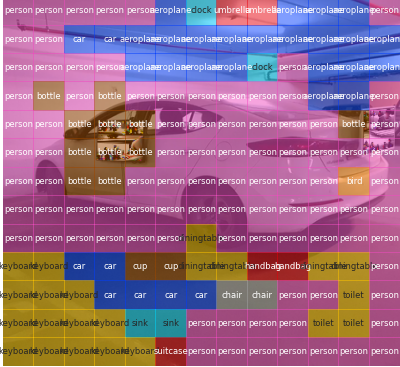}
    \caption{image with \textit{cls} loss patches}
    \label{fig:classes_heatmap_with_patches}
  \end{subfigure}
  \caption{Class map of non-patched image vs. class map of image with \textit{cls} loss patches.}\label{fig:classes_heatmap}
\end{figure}

On the other hand, we observe that the \textit{cls} loss patches do not harm the objectness score, as expected.
Figure \ref{fig:classes_heatmap} visualizes YOLO's classification output for the image.
The class mentioned in each cell is the class that obtained the maximum probability in the classes vector over all of the predictions for each cell.
In Figure \ref{fig:classes_heatmap_NO_patches}, we see the predicted class in each cell of a non-patched image, while Figure \ref{fig:classes_heatmap_with_patches} presents the same image with \textit{cls} loss patches (applied digitally).
The difference between the predicted classes in the two images over most of the cells is clear.
\textit{Cls} loss patches minimize the score of the class ``car" in the probability vectors, which causes the object to be classified incorrectly.
In this example, YOLO detects the car's boundaries, however, classifies it as ``person."

We were surprised to find that the \textit{obj\_cls} loss patches do not critically reduce the objectness scores (shown in Figure \ref{fig:obj_clsHeatmap}). 
Nevertheless, Tables \ref{table:leftRange_results} and \ref{table:2screens45Deg_results} show that this loss works very well most of the time.
Minimizing the \textit{obj\_cls} loss means minimizing the product score \((obj\_score*cls\_score)\), so we conclude that it is easier for the attack to reduce the \textit{cls} score, and by that, the \textit{obj\_cls} loss itself is minimized.

\subsection{Different Screen Sizes}
As there are screens of many different sizes and many possible locations to place patches on a car, we wanted to assess the impact of different screen sizes on the attack performance.
To do so, we trained patches of different sized screens and evaluated the attack in the digital world, i.e., placing the patch on each frame from the test video by using the placing algorithm and evaluating the attack success rate.
The training set consists of frames of a video of a car from a range of $-60\degree$  to $60\degree$  relative to the car's back, including viewpoints to both the left and right side.
This forms almost a semi-circle around the car. The video was filmed while just one screen was attached to the car's back.
The different sized patches were trained using the \textit{obj\_cls} loss.
In each evaluation we used one screen and one patch; thus, there was no patch switching based on the view angle in this experiment.\par
\begin{figure}[h]
  \centering
\includegraphics[scale=0.7]{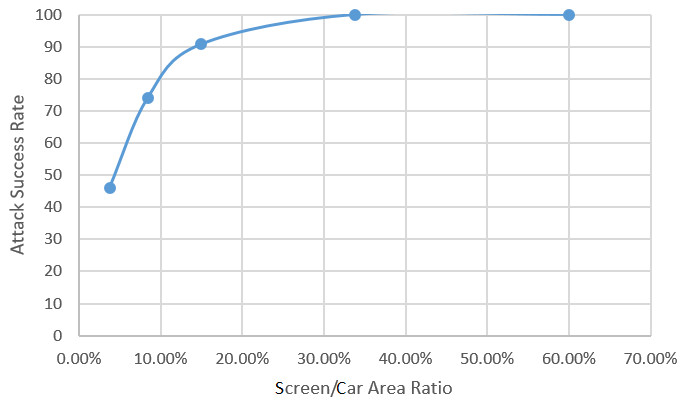}
  \caption{Attack success rate depending on \textit{screen/car area ratio}.}
  \label{fig:screenSizeGraph}
\end{figure}

We examined a few patch sizes relative to our 28'' screen and calculated the \textit{screen/car area ratio} - the ratio of the area that the screen covers to the area of the back of the car.
Our 28'' screen (used in the experiments of section \ref{sec_eval}) covers 15\% of the area at the back of the car.
Figure \ref{fig:screenSizeGraph} presents the results.
In the graph, it can be observed that while the original screen (which has a 15\% area ratio) achieves a success rate of over 90\%, smaller screen sizes obtain significantly lower rates.
Larger \textit{screen/car area ratio} screens obtain higher success rates than the original 28'' screen, however, the results differences aren't significant.
This actually describes the trade-off we observed - a larger screen size achieves better results, but the damage is much more serious in terms of the perturbation's size.
(We consider a high ratio as a damage for reasonable reasons; high ratio is caused by big screen covering large area of the car, hence the perturbation is much more noticeable).

\section{Transferability}
\subsection{Transferability Across Car Models}
An interesting issue to evaluate is whether the patches, which were trained on a dataset of a single car, are transferable across different car models (i.e., whether the attack's robustness is maintained).
To test this, we ran the experiment illustrated in Figure~\ref{fig:fullRangeAttack} again, however this time with two car models (\textit{Toyota Corolla 2006}, \textit{Fiat Tipo}) that differ from the training set car (\textit{Hyundai Elantra}).

The attack with the \textit{Fiat Tipo} achieved a success rate of 71\%.
The \textit{Toyota Corolla 2006} version had a success rate of 51\%.
These results can be explained by the fact that the appearance and color of the \textit{Fiat Tipo} are similar to that of the training set car.
In contrast, the color of the \textit{Toyota Corolla} is different than the training set car's color.
Therefore, we conclude that the attack is transferable among different car models, however it transfers better to cars which are similar to the training set car.

Furthermore, as shown in different papers (\cite{thys2019fooling, huang2020universal}), we assume that the robustness across different car models can be improved when training the patches over a training set that contains diverse samples of different car types and color.

\subsection{Transferability Across Object Detection Models}
We conducted another transferability experiment to examine the robustness of the attack across different object detection models.
The patches were originally trained to attack YOLOv2.
In this experiment, we tested the attack on other state-of-the-art object detection models: YOLOv3, Fast-RCNN, and Mask-RCNN (like the originally attacked detector, these models also were trained on the MS COCO dataset).
Similar to the experiment above, we assessed the transferability of the attack illustrated in Figure~\ref{fig:fullRangeAttack}.

During the evaluation, we discovered that the attack is not transferable to Fast-RCNN and Mask-RCNN, which have completely different architectures than YOLO.
Evaluating the attack on YOLOv3, which is built on a similar architecture to YOLOv2, we found that the \textit{cls}, \textit{obj}, and \textit{obj\_cls} loss achieved success rates of 13.8\%, 10.5\%, and 11.7\%, respectively.
Nevertheless, the attack range illustrated in Figure~\ref{fig:fullRangeAttack} can be divided into two sub-ranges (i.e., illustrated in Figures \ref{fig:2ScreensNumOfPatchesExp} and \ref{fig:NumOfPatchesSideExp}).
We analyzed the transferability results on each of the sub-ranges, and observed that while the attack did not work well in the view range of the car's side (Figure \ref{fig:NumOfPatchesSideExp}), it worked very well in the view range described in Figure \ref{fig:2ScreensNumOfPatchesExp}.
Hence, we conclude that the attack may transfer well to certain view angles when using models with similar architectures.

\subsection{Transferability Across Environments}
The experiments described in Section \ref{sec_eval} were conducted in an underground parking lot, where we couldn't control the environmental noise.
However, the area where the experiments took place resembled a cloudy day outside, where no direct sunlight was reflected on the screens.
Therefore, we conducted an experiment assessing the attack's transferability across different environments.
We performed the attack scenario illustrated in Figure~\ref{fig:fullRangeAttack}, however this time in an outside area at noon on a sunny day.
The attack success rates for each loss are presented in Table~\ref{table:envTransferabilityResults}.
The results show that the attack is ineffective in a sunny environment.
Comparing the results of the same attack performed in an underground parking lot (presented in Table~\ref{table:fullRangeCoverageResults}), we conclude that the sun significantly affects the attack effectiveness.
Strong sunlight increases the environment's brightness; it is also reflected on the screens, blinding the camera during filming, thus making it more difficult for the camera to catch the perturbations as computed, making them almost meaningless.

\begin{table}[h!]
\centering
\begin{tabular}{|c|c|c|c|} 
\cline{2-4}
    \multicolumn{1}{c|}{} & \textit{obj} loss & \textit{cls} loss & \textit{obj\_cls} loss \\ [0.5ex] 
 \hline 
    Success Rate & 15\% & 21.6\% & \textbf{23.2}\% \\ 
 \hline
\end{tabular}
\caption{Attack performance in a sunny environment, with a full quarter of a circle coverage.}
\label{table:envTransferabilityResults}
\end{table}

\section{Conclusion}
In this paper, we presented a novel adversarial attack method that aims to prevent object detectors from detecting large, nonplanar target objects.
The attack uses screens which are attached to the target objects' planes and leverages the screens' ability to switch between patches dynamically, according to the object detection system's position, in order to adjust the attack to the current situation and present the most effective patches.
The attack can be used in cases where the object detection system is static and the target object is in motion, (such as a car hiding from a speed camera) or vice versa.\par
We chose to demonstrate our method when the target object is a \textit{car}. We showed that the attack is effective against the state-of-the-art object detection system YOLOv2 in real-world settings.
In our experiments, we examined the optimal number of screens to attach to the car, the number of patches to present on the screens, the timing of the switches between patches, and the optimal loss function, for a wide range of view angles covering $90\degree$ (a full quarter of a circle around the car).

During the evaluations, we noticed that in some cases the attack caused the car's label to be changed to that of another vehicle, such as a bus or truck.
We addressed this by presenting the semantic adversarial patch, which forces the detector to switch the target object's label to a label that is not semantically-related to the original label.

We also tested the attack transferability and demonstrated that the attack can transfer to different car models, although it lacks transferability between different detection architectures.

We conducted a digital-world experiment to evaluate the effect of the screen size on the attack success and found that a screen covering 15\% of the back of the car is sufficient for fooling the detector in 90\% of the video frames.

To the best of our knowledge, this is the first study that leverages screens, and their advantages in such a situation, to present adversarial patches in the real world.

\section{Future Work}
There are many interesting research paths that can be taken following this study. First, transferability across environments can be improved.
Our experiments show that the attack is less effective on sunny days, when sunlight reflects on the screens.
We believe that using a different technique to present dynamic perturbations, such as LED lights attached to the car, may help in such a situation.
LED lights also can cover wide areas of the car (i.e., the roof, sides, etc.), in contrast to screens, which are usually limited to a square shape.\par
Second, our experiments show that the attack lacks transferability between models built on different architectures (such as YOLOv2 and Fast-RCNN).
Optimizing the objective function for different models simultaneously can improve the attack's robustness to different detection models.\par
Future research can also be performed on the semantic adversarial patch method, which now is handled manually; it is worth considering natural language processing techniques as a means of automatically choosing labels that are semantically-related to the original one.

\bibliographystyle{IEEEtranN}
\bibliography{ref}

\onecolumn
\appendix

\begin{algorithm}[h]
    \footnotesize
    \SetAlgoLined
    \SetKwInOut{Input}{Input}
    \SetKwInOut{Output}{Output}
    \SetKwFunction{genRndPatch}{GenerateRandomPatch}
    \SetKwFunction{rndTrans}{ApplyRandomTransformation}
    \SetKwFunction{placePatches}{PlacePatches}
    \SetKwFunction{detect}{DetectObjects}
    \SetKwFunction{computeLoss}{ComputeLoss}
    \SetKwFunction{totalVariation}{TotalVariation}
    \SetKwFunction{updatePatches}{UpdatePatches}
    \SetKwData{patches}{patches}
    \SetKwData{imgBatch}{img\_batch}
    \SetKwData{patchedBatch}{patched\_batch}
    \SetKwData{yoloOutput}{detection\_output}
    \SetKwData{loss}{loss}
    \SetKwData{totalTV}{total\_TV}
    \SetKwData{objective}{objective\_score}
    \Input{view\_angle\_dataset, \#\_of\_screens}
    \Output{optimized patches}
    
    \patches $\longleftarrow []$\;
    \For{$i\gets0$ \KwTo \#\_of\_screens}{
        $\patches[i] \longleftarrow$ \genRndPatch{}\;
    }
    \While{epoch $<=$ EPOCHS}{
        \For{\imgBatch $\in$ view\_angle\_dataset}{
            \totalTV $\longleftarrow 0$\;
            \rndTrans{\patches}\;
            \patchedBatch $\longleftarrow$ \placePatches{\imgBatch, \patches}\;
            \yoloOutput $\longleftarrow$ \detect{\patchedBatch}\;
            \loss $\longleftarrow$ \computeLoss{\yoloOutput}\;
            \totalTV $\longleftarrow$ \totalVariation{\patches}\;
            \objective $\longleftarrow \alpha*\totalTV + \loss$\;
            \patches $\longleftarrow$ \updatePatches{\objective}\;
        }
    }
    \Return \patches
      
    \caption{Patch Generation}
    \label{alg:PatchGeneration}
\end{algorithm}

\begin{algorithm*}[h]
   \footnotesize
    \SetAlgoLined
    \SetKwInOut{Input}{Input}
    \SetKwInOut{Output}{Output}
    \SetKwFunction{PatchGeneration}{PatchGeneration}
    \SetKwFunction{placePatches}{PlacePatches}
    \SetKwFunction{evaluateAttack}{EvaluateAttack}
    \SetKwFunction{splitDataset}{SplitDataset}
    \SetKwData{formerPatches}{former\_patches}
    \SetKwData{testImgs}{patched\_test\_imgs}
    \SetKwData{successRate}{former\_success\_rate}
    \SetKwData{newSuccessRate}{new\_success\_rate}
    \SetKwData{splitMore}{split\_more}
    \SetKwData{splittedDatasets}{train\_subsets}
    \SetKwData{splittedTestset}{test\_subsets}
    \SetKwData{splitTo}{split\_to}
    \SetKwData{subset}{train\_subset}
    \SetKwData{subsetsPatches}{subsets\_patches}
    
    \Input{train dataset, test dataset}
    \Output{patches and number of switches}
    \BlankLine
    
    \splitMore $\longleftarrow True$\;
    \splitTo $\longleftarrow 1$\;
    \subsetsPatches $\longleftarrow$ \PatchGeneration{train\_dataset, \#\_of\_screens}\;
    \testImgs $\longleftarrow$ \placePatches{test\_dataset, \subsetsPatches}\;
    \newSuccessRate $\longleftarrow$ \evaluateAttack{\testImgs}\;
    \BlankLine
    
    \While{\splitMore}{
        \splitTo$++$\;
        \testImgs $\longleftarrow []$\;
        \successRate $\longleftarrow$ \newSuccessRate\;
        \formerPatches $\longleftarrow$ \subsetsPatches\;
        \splittedDatasets $\longleftarrow$ \splitDataset{train\_dataset, \splitTo}\;
        \tcp{Compute patch for each subset:}
        \subsetsPatches $\longleftarrow []$\;
        \ForEach{$subset \in \splittedDatasets$}{
            \subsetsPatches.append(\PatchGeneration{subset, \#\_of\_screens})\;
        }
        \splittedTestset $\longleftarrow$ \splitDataset{test\_dataset, \splitTo}\;
        \tcp{Place the appropriate patch on each subset:}
        \For{$i\gets0$ \KwTo $\splitTo$}{
            \testImgs $+=$ \placePatches{$\splittedTestset[i], \subsetsPatches[i]$}
        }
        \newSuccessRate $\longleftarrow$ \evaluateAttack{\testImgs}\;
        \lIf{$\newSuccessRate <= \successRate$}{\splitMore = False}
    }
    \Return \formerPatches, $\splitTo - 1$

    \caption{Dynamic Attack}
    \label{alg:dynamicAttack}
\end{algorithm*}

\end{document}